\documentclass[aps,prl,superscriptaddress,singlecolumn]{revtex4}  
\usepackage{graphicx}  
\usepackage{dcolumn}   
\usepackage{bm}        
\usepackage{amssymb}   

\begin{document}



\title{A bremsstrahlung gamma-ray source based on stable \\ionization injection of electrons into a laser wakefield accelerator}
\author{A. D\"opp}
\email{andreas.doepp@polytechnique.edu}
\affiliation{LOA, ENSTA ParisTech, CNRS, \'Ecole polytechnique, Universit\'e Paris-Saclay, 828 bd des Mar\'echaux, 91762 Palaiseau Cedex, France}
\affiliation{Centro de Laseres Pulsados, Parque Cientfico, 37185 Villamayor, Salamanca, Spain}
\author{E. Guillaume}\affiliation{LOA, ENSTA ParisTech, CNRS, \'Ecole polytechnique, Universit\'e Paris-Saclay, 828 bd des Mar\'echaux, 91762 Palaiseau Cedex, France}
\author{C. Thaury}
\affiliation{LOA, ENSTA ParisTech, CNRS, \'Ecole polytechnique, Universit\'e Paris-Saclay, 828 bd des Mar\'echaux, 91762 Palaiseau Cedex, France}
\author{A. Lifschitz}
\affiliation{LOA, ENSTA ParisTech, CNRS, \'Ecole polytechnique, Universit\'e Paris-Saclay, 828 bd des Mar\'echaux, 91762 Palaiseau Cedex, France}
\author{F. Sylla}
\affiliation{SourceLAB SAS, 86 rue de Paris, 91400 Orsay, France}
\author{J-P. Goddet}
\affiliation{LOA, ENSTA ParisTech, CNRS, \'Ecole polytechnique, Universit\'e Paris-Saclay, 828 bd des Mar\'echaux, 91762 Palaiseau Cedex, France}
\author{A. Tafzi}
\affiliation{LOA, ENSTA ParisTech, CNRS, \'Ecole polytechnique, Universit\'e Paris-Saclay, 828 bd des Mar\'echaux, 91762 Palaiseau Cedex, France}
\author{G. Iaquanello}
\affiliation{LOA, ENSTA ParisTech, CNRS, \'Ecole polytechnique, Universit\'e Paris-Saclay, 828 bd des Mar\'echaux, 91762 Palaiseau Cedex, France}
\author{T. Lefrou}
\affiliation{LOA, ENSTA ParisTech, CNRS, \'Ecole polytechnique, Universit\'e Paris-Saclay, 828 bd des Mar\'echaux, 91762 Palaiseau Cedex, France}
\author{P. Rousseau}
\affiliation{LOA, ENSTA ParisTech, CNRS, \'Ecole polytechnique, Universit\'e Paris-Saclay, 828 bd des Mar\'echaux, 91762 Palaiseau Cedex, France}
\author{E. Conejero}
\affiliation{Departamento de F\'isica Aplicada, Universidad de Salamanca, Plaza de la
Merced s/n, 37008 Salamanca, Spain}
\author{C. Ruiz}
\affiliation{Departamento de F\'isica Aplicada, Universidad de Salamanca, Plaza de la
Merced s/n, 37008 Salamanca, Spain}
\author{K. Ta Phuoc}
\affiliation{LOA, ENSTA ParisTech, CNRS, \'Ecole polytechnique, Universit\'e Paris-Saclay, 828 bd des Mar\'echaux, 91762 Palaiseau Cedex, France}
\author{V. Malka}
\affiliation{LOA, ENSTA ParisTech, CNRS, \'Ecole polytechnique, Universit\'e Paris-Saclay, 828 bd des Mar\'echaux, 91762 Palaiseau Cedex, France}

\begin{abstract}
Laser wakefield acceleration permits the generation of ultra-short, high-brightness relativistic electron beams on a millimeter scale. While those features are of interest for many applications, the source remains constraint by the poor stability of the electron injection process. Here we present results on injection and acceleration of electrons in pure nitrogen and argon. We observe stable, continuous ionization-induced injection of electrons into the wakefield for laser powers exceeding a threshold of 7 TW. The beam charge scales approximately linear with the laser energy and is limited by beam loading. For 40 TW laser pulses we measure a maximum charge of almost $1$ nC per shot, originating mostly from electrons of less than 10 MeV energy. 
The relatively low energy, the high charge and its stability make this source well-suited for applications such as non-destructive testing. Hence, we demonstrate the production of energetic radiation via bremsstrahlung conversion at 1 Hz repetition rate. In accordance with \textsc{Geant4} Monte-Carlo simulations, we measure a $\gamma$-ray source size of less than $100$ microns for a 0.5 mm tantalum converter placed at 2 mm from the accelerator exit. Furthermore we present radiographs of image quality indicators.
\end{abstract}

\maketitle

\section{Introduction}

Since the first proposal in the late 1970s \cite{Tajima:1979un}, laser wakefield accelerators have gone a long way from a theoretical concept to a reliable source of highly relativistic electrons. While mostly known for its compactness, resulting from the gigavolt to teravolt per meter field gradients inside the plasma cavity \cite{Esarey:2009ks}, this type of accelerator also inherently provides beams of femtosecond duration \cite{Lundh:2011js} and micrometer diameter \cite{Kneip:2012gx}. Driven by ambitious goals like table-top free electron lasers \cite{Nakajima:2008cs}, many efforts have been dedicated to improvements of the transverse emittance \cite{Plateau:2012he} and the energy spread \cite{Faure:2006vy} as well. However, these developments usually result in an increased experimental complexity and a reduced beam charge of a few picocoulomb per shot.

In contrast, temporally incoherent radiation sources are less constraint in terms of beam quality and work well at high beam charge ($>100$ pC). Most prominent examples are the synchrotron-like betatron \cite{Rousse:2004tc,Kneip:2010kk} and Compton \cite{TaPhuoc:2012cg,Schwoerer:2006dw} sources. It has been shown that these sources can be used for single-shot X-ray imaging \cite{Fourmaux:2011cs,Dopp:2015tl}, yet their robustness is still not sufficient to compete with conventional solutions. A more simple mechanism is to create high energy radiation via bremsstrahlung emission in a high-Z material. This technique, analogous to conventional X-ray tubes \cite{Coolidge:1913wr}, was first demonstrated in 2002 \cite{Edwards:2002cl} and subsequent experiments have demonstrated that the source is suitable for high resolution imaging in non-destructive testing \cite{Glinec:2005ve,BenIsmail:2011uq}.

However, the electrons used in these experiments originated from spontaneous self-injection into the wake \cite{Corde:2013gj} and typically reached energies in the order of 100 MeV. Not only is this kind of electron injection very unstable and therefore unsuitable for many applications, but furthermore it is desirable to operate at energies below 10 MeV. The reason for this is that significant neutron contamination occurs at higher energies \cite{Chen:2006ho} and such a source would then require additional radioprotection \cite{Soriani:2010id}.

It is the injection method which determines many source parameters such as the energy spread, charge and stability. The aforementioned self-injection is the most common injection method and it was shown that this process can lead to the production of quasi-monoenergetic electron beams  \cite{Faure:2004tj,Geddes:2004vs,Mangles:2004vr}. Unfortunately its spontaneous nature results in an unreliable performance and different controlled injection schemes have been developed to address this issue, including heating of electrons with a colliding laser pulse \cite{Faure:2006vy,Malka:2008fm} or controlled cavity expansion in a density downramp \cite{Geddes:2008tj,Schmid:2010ih}. Here we have employed the ionization injection method \cite{McGuffey:2010wy}, which is well-suited to fulfill the above requirements of the source concerning stability and beam energy. As we will discuss in the following sections, ionization-induced injection can provide highly charged electron beams and we use short jets of pure high-Z gases in order to reduce the beam energy.

\section{Ionization-induced injection}
Laser wakefield accelerators usually operate at peak laser intensities in excess of $10^{18}$ W.cm$^{-2}$. At such intensities already the leading edge of the laser pulse can entirely ionize gas targets consisting of hydrogen or helium. The situation changes when high Z gases like nitrogen (N), carbon-dioxyde (CO$_2$) or argon (Ar) are employed. Here the outer shells, whose binding energies are typically below 100 eV, are likewisely ionized at the very front of the laser pulse. However, higher ionization states such as Ar$^{9+}$ and N$^{7+}$ will only be reached close to the peak of the pulse and are therefore ionized with a delay. This is illustrated in figure 1, which shows both ionization and wakefield excitation for a laser pulse propagating through argon. The driver is modeled as a $\sin^2$ shaped 800 nm pulse with a peak amplitude of $a_0=1.0$ at a full width at half maximum (FWHM) of half a plasma wavelength $\lambda_p$ at a density of $n_e=10^{19}$cm$^{-3}$ ($\lambda_p/c_0\sim$ 35 fs). The wakefields and possible electron trajectories are calculated using the one-dimensional wakefield model \cite{Esarey:1995uh}. Trapped - and therefore accelerated - electron orbits are plotted in solid green lines, while non-trapped trajectories are dashed. The injection threshold - the separatrix - is marked in black. Ionization rates are calculated using the ADK tunneling ionization model \cite{Delone:1998gw} and the early (late) ionization region is marked with a yellow (red) shaded rectangle. As a result of their late ionization, the inner shell electrons experience asymmetric longitudinal wakefields, meaning that they can gain a signifiant amount of longitudinal momentum in direction of the propagation and therefore trapping into the wake is facilitated \cite{Oz:2007tz,Chen:2012du}. 


\begin{figure}[t]\centering
\includegraphics[width=0.85\linewidth]{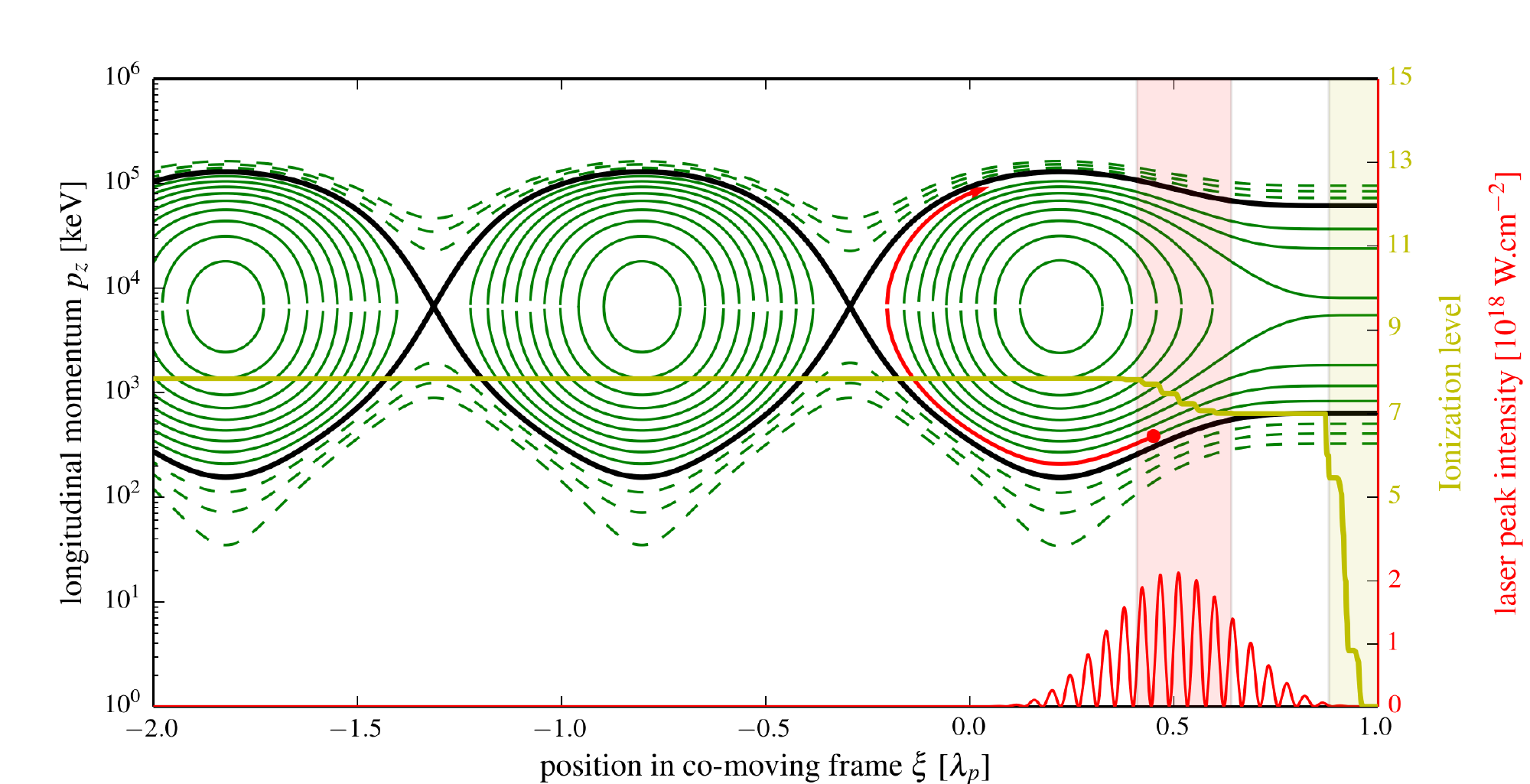}
\caption{Illustration of ionization-induced injection in Argon. The plasma has a density of $n_e=10^{19}$cm$^{-3}$ and the laser (red) has a peak amplitude of $a_0=1.0$ at a full width at half maximum of half a plasma wavelength $\lambda_p$ ($\sim$ 17.5 fs). Trapped electron orbits are shown in solid green lines, while non-trapped trajectories are dashed. The separatrix is marked in black. Also shown is the ionization of argon according to the ADK ionization model (yellow solid line). In this representation is becomes clear that the early ionized outer shell electrons (1-7, yellow box) need more energy to get trapped than the higher ionization states (red box).}
\label{fig1}
\end{figure}

As a consequence of this, a laser plasma accelerator based on ionization-induced injection can be operated at lower plasma densities than accelerators relying on self-injection. This can mitigate effects like electron dephasing or laser depletion, and consequently accelerators using ionization-induced injection have shown to lead to higher electron energies \cite{Clayton:2010kh}. However, the ionization (and thus injection) occurs continuously during the laser propagation, which is why ionization-induced injection usually leads to broad energy spectra, see for example \cite{McGuffey:2010wy}. Once a significant amount of electrons are trapped, their proper fields counteract the wakefields and this beam loading limits the maximum charge which is attainable in the accelerator.

Most studies on ionization-induced injection rely on helium-dominated gas mixtures, with only a few percent of the high-Z gas. The laser propagation is then similar to pure helium. In contrast, this study has been performed using pure nitrogen and argon. In this regime the laser propagation is affected by ionization-induced defocusing, as seen in the shadowgraphy (Fig.\ref{fig2}b). In the next section we will discuss the performance of the accelerator in this configuration. More details on electron acceleration in this regime have been published elsewhere \cite{Guillaume:2015gy}.

\section{The laser plasma accelerator}

\begin{figure}\centering
\includegraphics[width=0.97\linewidth]{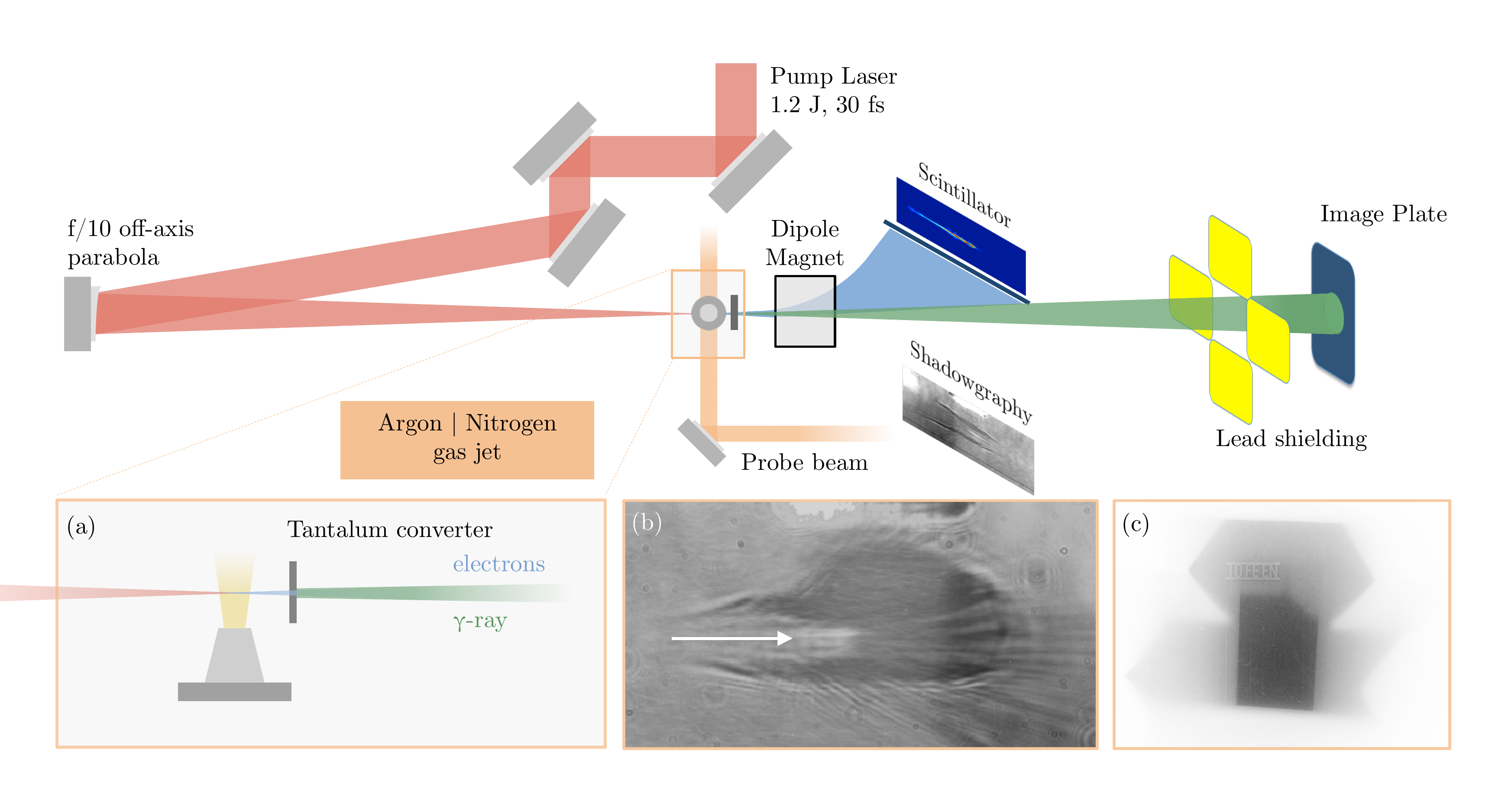}
\caption{Schematic setup of the experiment. The multi-TW laser pulse is focussed into a gas jet of nitrogen or argon. Inner shell electrons are injected via delayed tunneling ionization into the wake of the pulse and accelerated. Once they exit the gas jet they penetrate a tantalum foil, leading to the emission of bremsstrahlung, cf. inlet (a). This bremsstrahlung (green) is then detected on an image plate, while the charge and energy of the electron beam (blue) is measured in an absolutely calibrated spectrometer. (b) shows a typical shadowgraphy image of the plasma channel  created by the laser. (c) shows a cropped image plate scan, where regions of highest radiation exposure have turned dark. The triangular shape around the object is the shadow cast by the lead shielding. }
\label{fig2}
\end{figure}

The experimental setup is shown in figure \ref{fig2}. As driver of the accelerator we use the \textsc{Salle Jaune} Ti:Sa laser system, which delivers 28 fs pulses of up to 2.1 joule energy at a central wavelength of $\lambda_0=800$ nm. The 70 mm diameter flat-top laser pulses are focused on a gas target using a 690 mm off-axis parabolic mirror. Aberrations are corrected with a deformable mirror, finally leading to an Airy-like focal spot with a central mode of 22 $\mu$m 1/e$^2$ diameter that contains 52 percent of the total beam energy. The beam energy can be tuned using a $\lambda/2$ plate followed by a polarizer. For the experiment we scanned over peak intensities of $1.6 \times 10^{18}$ W.cm$^{-2}$ to $8.9 \times 10^{18}$ W.cm$^{-2}$.

The gas target consists of a parker series 9 valve onto which an exit nozzle of 0.7 mm diameter is mounted, connected to a reservoir of either argon or nitrogen. The valve opens 10 milliseconds before the laser pulse arrives, letting the gas expand sonically into the vacuum. At the laser focus (0.4 mm above the exit of the jet) the gas profile extends over a length of 1.4 mm.

The plasma electron density is measured with a probe beam using a Nomarski interferometer \cite{Benattar:1979uha}. The electron energy distribution and charge are measured via dispersion of electrons through a dipole magnet. The electrons are detected using a Kodak Lanex phosphor screen, imaged onto a 16-bit CCD camera.


\begin{figure}\centering
\includegraphics[width=1.0\linewidth ]{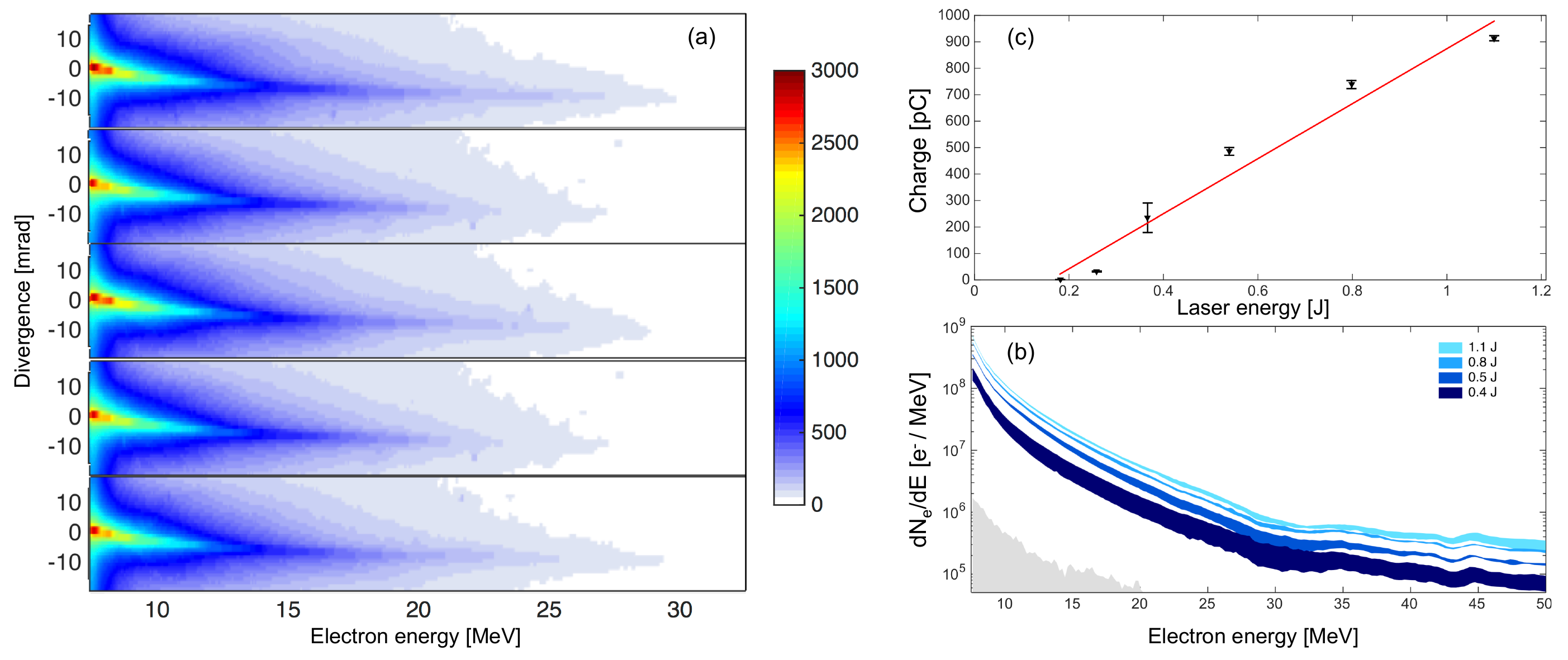}
\caption{Frame (a): Angularly resolved electron spectra for 5 consecutive shots. The accelerator operates extraordinary stable in this regime. (b) Average beam spectra for various beam energies, The line width corresponds to the error, showing the high stability of the source. (c) Scaling of the integrated beam charge with laser energy.}
\label{fig3}
\end{figure}


Due to the continuous injection, the electron spectra are not monoenergetic, but rather thermal. Below 10 MeV the distribution has a temperature of around $2$ MeV, while the temperature between 10 and 30 MeV is about $4$ MeV. This rather low beam energy is a consequence of plasma defocusing and the short length of the jet.

The beam divergence is energy dependent. For the 'cold' part of the spectrum the divergence reaches up to 20 mrad, while the divergence above 10 MeV is about 10 mrad (FWHM). With argon as target gas we observe an injection threshold of 0.2 J at an ion density of $\sim 2.4\times 10^{18}$cm$^{-3}$. From there on we observe that the beam charge increases linearly by 110 pC/100 mJ, leading to 910 pC maximum beam charge above 7 MeV at full laser energy (1.1 joule). The electron source is remarkably stable, both in terms of charge and energy. Furthermore the accelerator is less sensitive to the focal spot quality than in the self-injection regime.


\section{Conversion to bremsstrahlung}

Because of its superior stability and robustness the source is well suited for applications. Here we have applied it for the production of X/$\gamma$-rays via conversion into bremsstrahlung. In contrast to the preceding studies of this type, which relied on self-injection \cite{Glinec:2005ve,BenIsmail:2011uq}, using ionization-induced injection we could provide a stable electron source that was operated at the nominal laser repetition rate of 1 Hz.

Just as in a conventional X-ray tube, radiation is generated via penetration of a solid target with electrons. For the convertion we use tantalum foils of different thicknesses (0.5 mm, 1.0 mm and 1.5 mm) in order to study their influence on the source size. Using the continuous-slowing-down approximation we estimate a stopping range of $0.5$ mm for 1 MeV electrons, which means that they will slow down to rest within the converter. Though most of the stopping power goes into coulomb collisions, about $10\:\%$ of electron energy is converted into radiation. At 10 MeV the stopping range increases to $3.7$ mm, meaning that these electrons are not stopped within the converter. However, the ratio of radiative stopping power to collision stopping power increases significantly between 1 and 10 MeV, and accordingly we estimate a radiation yield of around $6-20$ percent for targets of $0.5-1.5$ millimeter thickness. As a first rough estimation we expect the production of $\sim10^{-4}$ Joule of radiation per shot. 

As illustrated in figure 2 the converter foils are placed behind the gas jet, at distances between 5 and 20 millimeters. The X-ray signal is measured using photostimulable phosphor plates (Fuiji BAS TR). The response of these image plates depends essentially on the energy deposited in the phosphor layer and therefore drops significantly for photon energies above 100 keV \cite{Meadowcroft:2008fo}. We estimate the photon spectrum using \textsc{Geant4} \cite{Agostinelli:2003iv} simulations, which is shown in figure 4.

\begin{figure}\centering
\includegraphics[width=0.8\linewidth,trim=0 3cm 0 2cm ,clip]{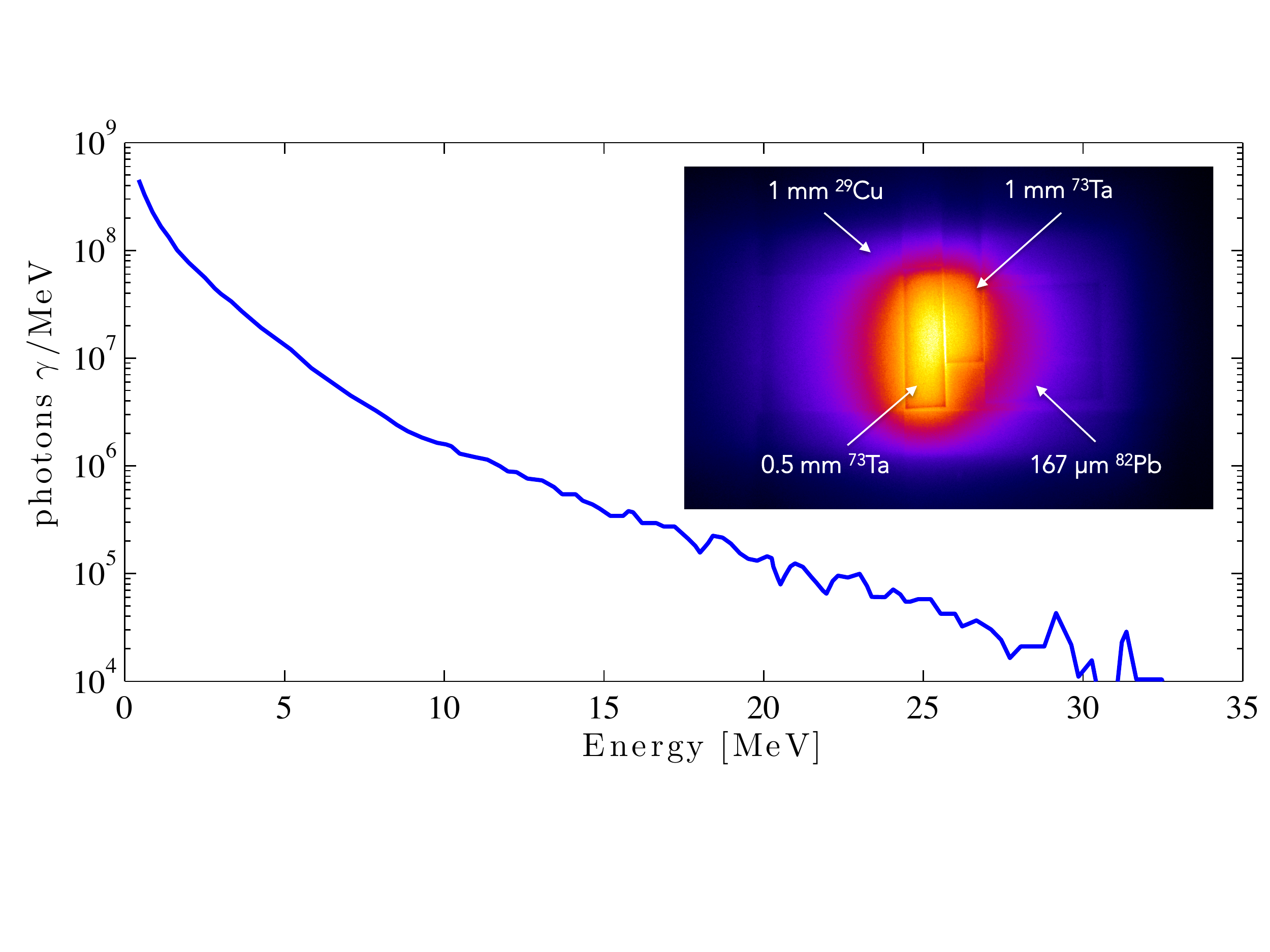}
\caption{$\gamma$-ray spectrum retrieved from \textsc{Geant4} simulations. Inset: Absorption contrast of different filters.}
\label{fig4}
\includegraphics[width=0.8\linewidth,trim=0 3cm 0 2cm ,clip]{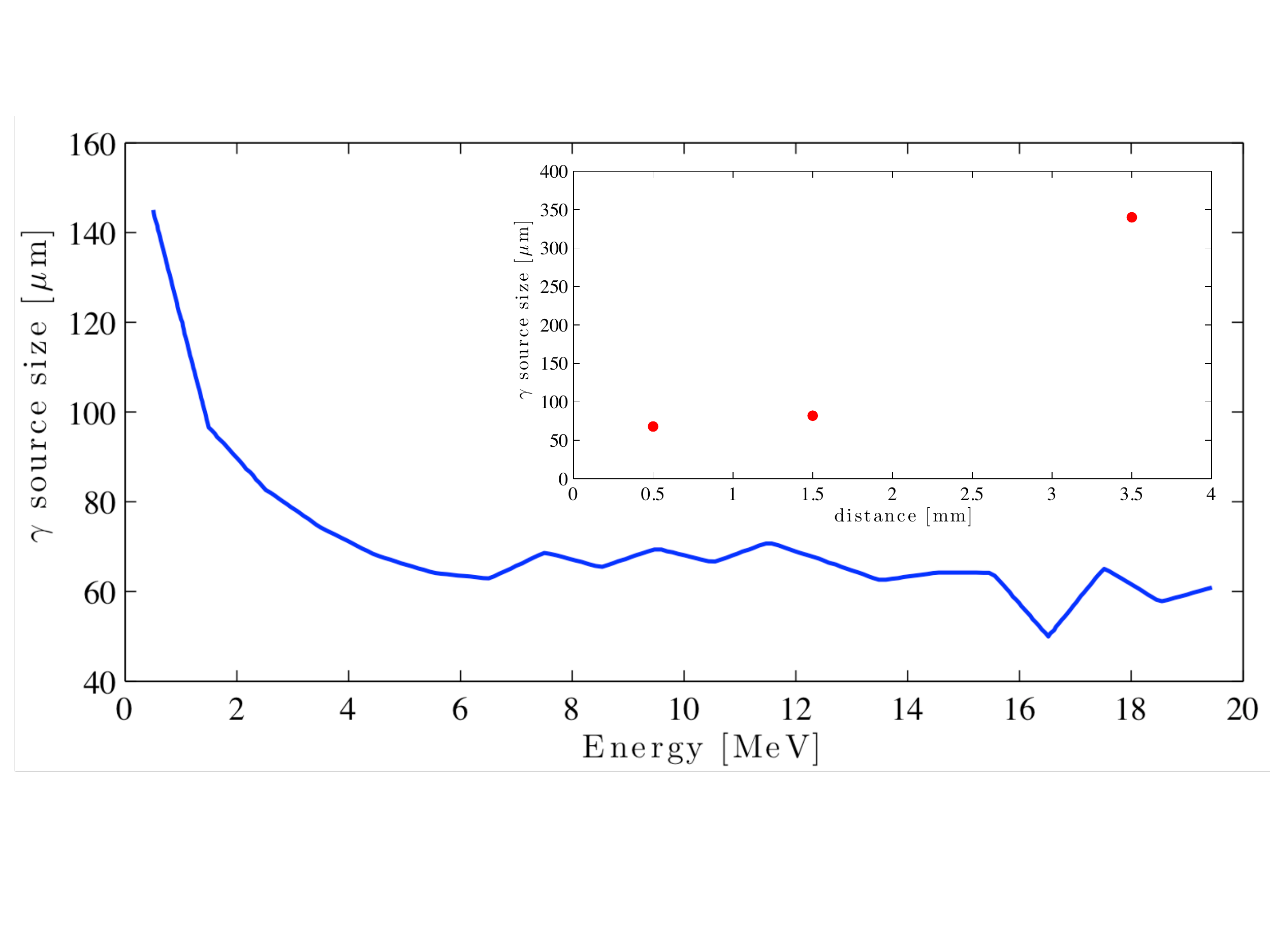}
\caption{ \textsc{Geant4} simulation of the source size at 2mm from the jet. Inset: Knife edge measurements for different distances.}
\label{fig1}
\includegraphics[width=0.75\linewidth,trim=3cm 5cm 3cm 5cm,clip ]{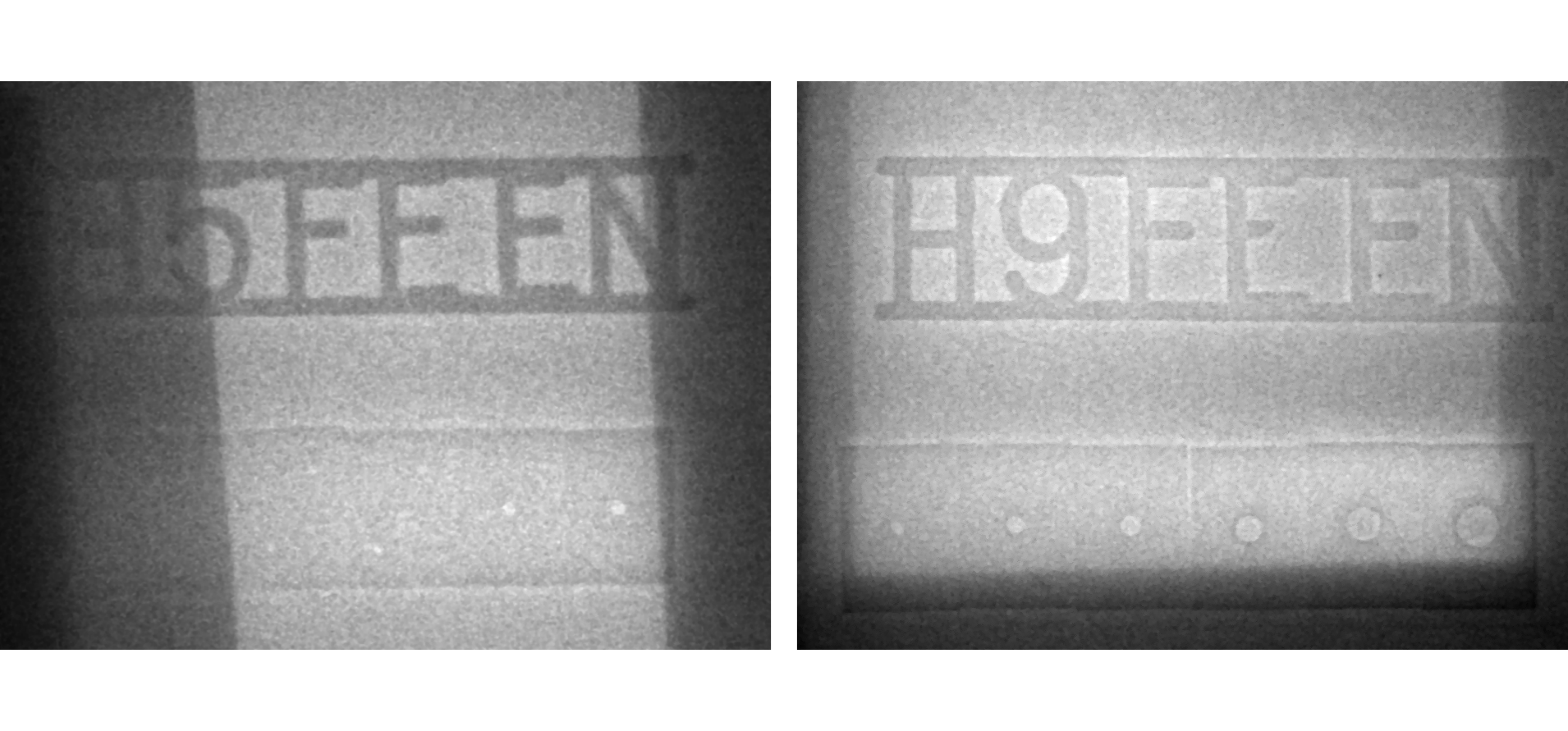}
\caption{Images of step-hole image quality indicators. Dark areas are where part of the lead shielding is in front of the image plate.}
\label{fig6}
\end{figure}

Using a knife-edge we have measured the X-ray source size for different distances between the gas jet exit and the converter. At $0.5$ mm from the exit the source size is estimated to be $65$ $\mu$m and no significant influence of the converter thickness (0.5 mm,1 mm and 1.5 mm) is observed. As shown in figure 5 the source size remaines below $100$ microns at 1.5 mm distance and at 3.5 mm from the gas jet it augments to $350$ $\mu$m. While the results are roughly in accordance with Monte-Carlo simulations of the source size for a 0.5 mm tantalum converter placed at $2$ millimeters from the exit, it should be noted that the source size depends strongly on the properties of the scattered electrons. As discussed above, we expect low energy electrons to undergo strong scattering, up to complete stopping inside the converter. Furthermore the initial beam divergence is about an order of magnitude larger for low energy electrons than it is for electrons with energies above 10 MeV. This tendency is reproduced in simulations (see figure 5), where we see that the $\gamma$ source size increases significantly at lower energies.

During the experiment we initially noticed a very low image contrast, due to a high background noise level. It is found that this noise originates from electrons hitting the chamber wall and the image quality is significantly improved with additional lead shielding. Still, radiographic applications are limited by the relatively bad signal-to-local-noise ratio $\sigma_S/\langle S \rangle\sim 0.1$, where the local noise $\sigma_S$ is the standard deviation of the signal in the region taken for calculation of the average signal $\langle S \rangle$. While this prevents imaging of weakly absorbing objects, we are able to perform some radiographies in order to assess the suitability of the source for imaging applications. As an example, figure 6 shows radiographies of industry standard (DIN EN 462) image quality indicators. The smallest features resolved have a size in the order of 200 micrometers.

\section{Conclusions}

In conclusion we have demonstrated a bremsstrahlung $\gamma$-ray source that relies on a stably operating laser-plasma accelerator. Its performance is a result of the ionization injection mechanism in pure argon and nitrogen, which leads to continuous electron injection starting from a threshold of 200 mJ pulse energy ($a_0\sim 0.8$). The electron beams have a quasi-Maxwellian spectrum and exhibit a high beam charge of almost 1 nC at 1 J pulse energy. The shot-to-shot stability is very good for a laser plasma accelerator and the source was in permanent 1 Hz operation over hundreds of shots. Using a tantalum converter, we have demonstrated the production of gamma radiation with less than 100 micrometer source size and features of 200 micrometer size can be resolved on radiographs of image quality indicators. We have deduced the radiation spectrum using \textsc{Geant4} simulations and show that most of the radiation is less energetic than 10 MeV, which is important for radioprotection. In conjunction with next generation high-repetition rate laser systems, this configuration could soon serve as competitive radiation source for applications such as non-destructive imaging.

\section*{Acknowledgements}
This work was supported by the European Research Council through the PARIS ERC project (Contract No. 226424), the X-Five ERC project (Contract No. 339128), LA3NET (Grant Agreement No. GA-ITN-2011-289191), Areva NDE, the NANOBIODOS INCA project and by the Agence Nationale pour la Recherche through the projects ANR-10-EQPX-CILEX and FENICS ANR-12-JS04-0004-01. We acknowledge Areva NDE Solutions for providing image quality indicators.




\bibliography{bremsstrahlung.bib}

\end{document}